\documentclass[aps,prb,twocolumn,groupedaddress,showpacs]{revtex4}

\usepackage{graphicx}

\begin{document}


  \title{Quantum corrections to the conductivity and Hall coefficient of a
2D electron gas in a dirty AlGaAs/GaAs/AlGaAs quantum well:
transition from diffusive to ballistic regime.}



  \author{V.~T.~Renard$^{1,3}$, O.~A.~Tkachenko$^{2,1}$, Z.~D.~Kvon$^{2,1}$,
E.~B.~Olshanetsky$^{2,1}$, A. I. Toropov$^2$, J~.C.~Portal$^{1,3,4}$}
\affiliation{$^1$ GHMFL, MPI-FKF/CNRS, BP-166, F-38042, Grenoble
Cedex9, France;} \affiliation{$^2$ Institute of Semiconductor
Physics, Novosibirsk 630090, Russia;}\affiliation{$^3$
INSA-Toulouse, 31077, Cedex 4, France;}\affiliation{$^4$ Institut
Universitaire de France, Toulouse, France;}


\date{15 December 2004}

  \begin{abstract} We report an experimental study of the quantum
corrections to the longitudinal conductivity and the Hall
coefficient of a low mobility, high density two-dimensional
two-dimensional electron gas in a AlGaAs/GaAs/AlGaAs quantum well
in a wide temperature range (1.5 K - 110 K). This temperature
range covers both the diffusive and the ballistic interaction
regimes for our samples. It was therefore possible to study the
crossover region for the longitudinal conductivity and the Hall
effect.
\end{abstract}


\maketitle

\section{Introduction}
  At low temperatures the conductivity of a degenerated two-dimensional
electron gas (2DEG) is governed by quantum corrections to the
Drude conductivity $\sigma_D$. In general, these corrections have
two principal origins: the weak localization (WL) and the
electron-electron (e-e) interaction \cite{Lee}. Until recently our
understanding of the interaction corrections to the conductivity
of a 2DEG was based on the seemingly unrelated theories developed
for two opposite regimes: the diffusive regime\cite{Altshuler}
$k_BT\tau/\hbar\ll1$, and the ballistic regime\cite{Gold}
$k_BT\tau/\hbar\gg1$. In the diffusive regime the quasi-particle
interaction time $\hbar/k_BT$ is larger than the momentum
relaxation time $\tau$ and two interacting electrons experience
multiple impurity scattering. In the ballistic regime the e-e
interaction is mediated by a single impurity. Recently, Zala,
Narozhny, and Aleiner (ZNA) have developed a new theory of the
interaction related corrections to the conductivity
\cite{Zala,Zala1} that bridges the gap between the two theories
known previously \cite{Altshuler,Gold}. One of the important
conclusions of the new theory is that the interaction corrections
to the conductivity in both regimes have a common origin: the
coherent scattering of electrons by Freidel oscillations.
Conformably to the previous results \cite{Altshuler,Gold}, the new
theory predicts a logarithmic temperature dependence of the
longitudinal conductivity and the Hall coefficient in the
diffusive regime, whereas in the ballistic regime the temperature
dependence of these parameters becomes linear and $T^{-1}$
respectively.

 Despite a surge of experimental activity
\cite{Coleridge,Shashkin,Proskuryakov,Kvon,Olshanetsky,Yasin}
following the publication of the theory \cite{Zala,Zala1} so far
no experiment has been reported where the transition between the
two regimes would have been clearly observed. One of the reasons
is that the temperature at which the transition is expected to
occur is given by $k_BT\tau/\hbar\approx0.1$, so that in the
relatively high-mobility 2D systems that are commonly studied the
transition temperature is by far too low ($T<100$ mK for
$\tau>10^{-11}$ sec). Thus, the ZNA theory has so far been
verified only in the intermediate and ballistic regimes
\cite{Galaktionov} ($k_BT\tau/\hbar=0.1-10$).

 To shift the transition to higher temperatures one should use low
mobility samples (small $\tau$). At the same time high carrier
densities $N_s$ are necessary in order to maintain high
conductivity and avoid strong localization. Moreover in high
density 2D systems the characteristic parameter
$r_s=E_C/E_F\propto1/N_s^{1/2}$, the ratio between Coulomb energy
and Fermi energy is small ($r_s<1$) and hence the effect of e-e
interaction is relatively weak. In this case the Fermi liquid
interaction constant $F_0^\sigma$, the only parameter in the
expressions for the quantum corrections to the conductivity in the
theory \cite{Zala}, can be calculated explicitly.

 In this respect low-mobility high-density systems
appear to offer certain advantages for testing the theory
\cite{Zala,Zala1}, as compared to high-mobility low-density
systems. Indeed not only they provide an opportunity for studying
an experimentally accessible temperature transition between the
diffusive and the ballistic interaction regimes but also the
comparison between the theory and experiment requires no fitting
parameters.

  The aim of the present work is to experimentally study the interaction
related corrections to the conductivity and the Hall coefficient
in a broad temperature range covering both the diffusive and
ballistic interaction regimes and the transition between them. The
experimental results obtained in the weak interaction limit are
expected to allow for a parameter free comparison with the ZNA
theory.

\section{Experimental set up}

  The experimental samples had a 2DEG formed in a narrow ($5$~nm)
AlGaAs/GaAs/AlGaAs quantum well $\delta$-doped in the middle. Such
doping results in a low mobility and a high carrier density. A
detailed description of the structure can be found in
Ref.~\onlinecite{Kvon1}. Two samples from the same wafer have been
studied for which similar results were obtained. Here we present
the data obtained for one of the samples with the following
parameters at $T=1.4$ K depending on prior illumination: the
electron density $N_s=(2.54-3.41)\times10^{12}$ cm$^{-2}$ and the
mobility $\mu=(380-560)$ cm$^2$/Vs. The Hall bar shaped samples
were studied between 1.4 K and 110 K in magnetic fields up to 15 T
using a superconducting magnet and a VTI cryostat and also a flow
cryostat ($T>5$~K) placed in a 20 T resistive magnet. The data was
acquired via a standard four-terminal lock-in technique with the
current $10$~nA.

\begin{figure}
\includegraphics[angle=-90,width=\columnwidth]{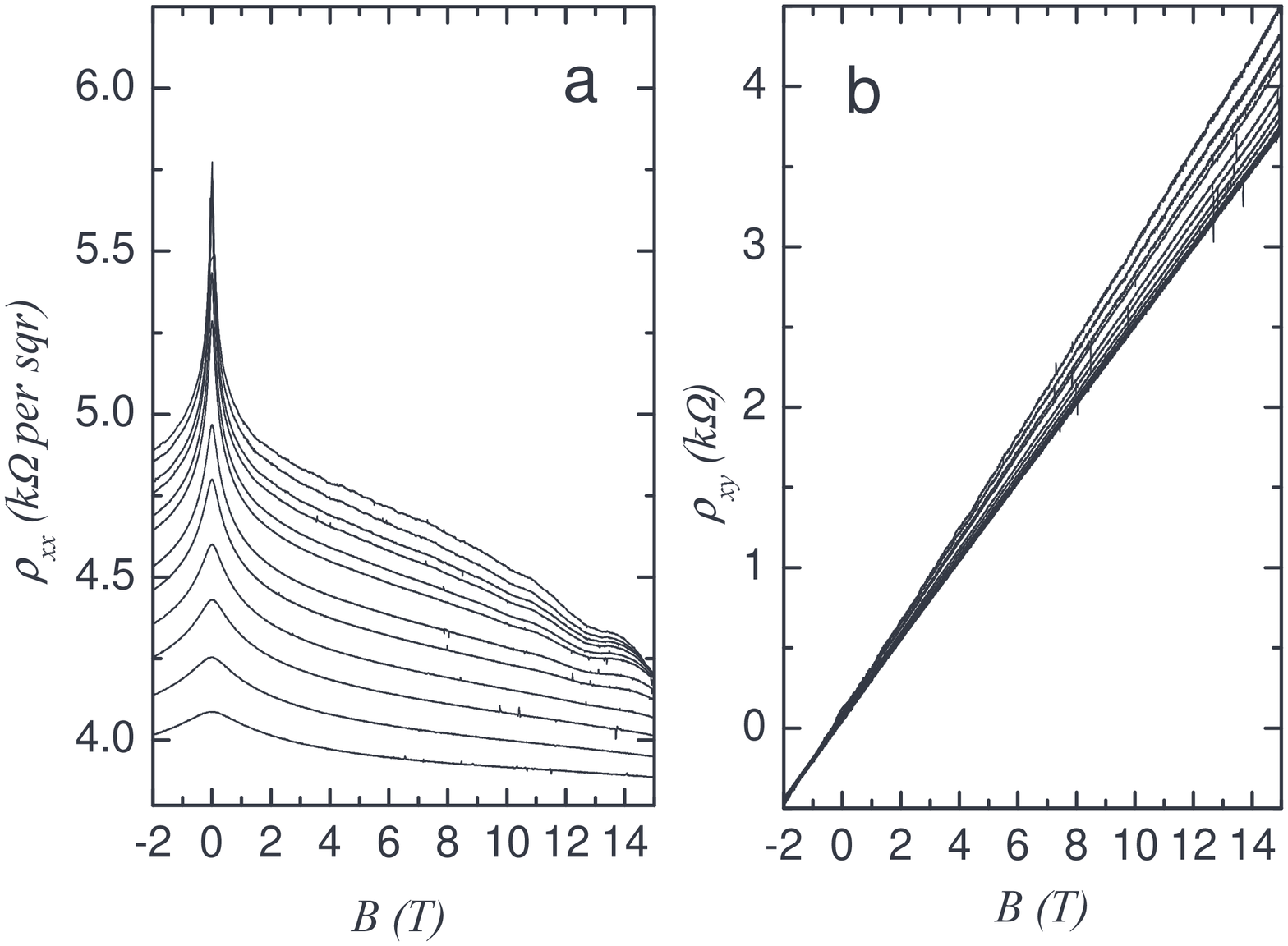}%
\caption{\label{Figure1} a) Longitudinal resistivity of the sample
at $N_s=2.56\times10^{12}$ cm$^{-2}$ for temperature=1.4 K, 1.9 K,
3.1 K, 4 K, 7.2 K, 10.25 K, 15.45 K, 21.5 K, 31 K, 46.2 K, 62.8 K,
84.5 K and 110 K from top to bottom. b) Hall resistance at the
same temperatures (from top to bottom).}
\end{figure}

  Fig.~\ref{Figure1} shows the longitudinal and Hall resistances of the
sample as a function of magnetic field at temperatures up to
110~K. As can be seen both are strongly temperature dependent.
Before analyzing the role of the quantum corrections in the
behavior of the transport coefficients shown in
Fig.~\ref{Figure1}, let us estimate the possible contribution from
other unrelated temperature dependent factors.

  First, since the measurements were performed up to relatively high
temperatures, the question of the role of phonon scattering
becomes important. In this connection we believe that the
following argument can be used. It is well known that in
ultra-clean GaAs samples sufficiently high values of mobility are
reported even at liquid nitrogen temperatures (see, for example
Ref.~\onlinecite{Pfeiffer,Lin}, where $\mu=4\times10^5 cm^2/Vs$ at
$T=77$~K). At these temperature the phonon scattering is the
dominant scattering mechanism in these samples and yet the
mobilities are still a thousand times larger than in our sample.
In our experiment, the phonon contribution to the conductivity
(and thus its variation with temperature) is thus expected to be
around 0.1\% at the highest temperature and can be neglected in
the entire temperature range.

  Now, as can be seen in Fig.~\ref{Figure1}, the slope of the Hall
resistance versus $B$ dependence varies with $T$ at low
temperatures but remains practically constant for $T>20$~K. One
might argue that the behavior at low temperatures could be due to
a variation of the electron density with temperature. However, we
believe that this is not the case. Indeed, from the measurements
carried out up to 20 T where the Shubnikov de Haas oscillations
are better resolved, we find that the density remains constant at
$T<30$~K. Also we find that the density given by the SdH
oscillations is the same as we get from the slope of the Hall
resistance at $T>20$~K where it is $T$-independent. We conclude
therefore that the electron density remains constant in the entire
experimental temperature range and all the data presented in
Fig.~\ref{Figure1} corresponds to $N_s=2.56\times10^{12}$
cm$^{-2}$.

  Having excluded the phonon scattering and the density variation as
possible causes of the behavior shown in Fig.~\ref{Figure1} we
associate the observed temperature dependences with the quantum
corrections to the transport coefficients. Our data will be
analyzed in the framework of the recent theories \cite{Zala,Zala1}
valid for a degenerated 2DEG ($k_BT\ll E_F$). According to
Ref.~\onlinecite{Kvon1} only one subband is occupied in our
quantum wells at $N_s=2.56\times10^{12}$ cm$^{-2}$. Also
$E_F\approx1000$~K and so the theory \cite{Zala,Zala1} should
apply under our experimental conditions.

\section{Longitudinal conductivity at B=0~T}

 Let us first describe how the experimental quantum corrections
were extracted from the row data and then turn to the analysis of
the obtained corrections.

  With the magnetic field increasing the MR in Fig.~\ref{Figure1}a goes
through two distinct types of behavior. An abrupt drop of
resistance at low fields and then a much weaker magnetic field
dependence at higher $B$. It is easy to show that the possible
classical MR described in Ref.~\onlinecite{Dmitriev1} can be
neglected in our sample. Indeed, the fraction of circling
electrons, which in this theory are supposed to cause a deviation
from the Drude theory is very small in our samples due to the low
electron mobility. This means that the behavior in
Fig.~\ref{Figure1}a must be attributed to quantum interference
effects, such as WL and the electron-electron interaction related
corrections to the conductivity. As is well known the weak
localization is suppressed at magnetic fields larger than
$B_{tr}=\hbar/(2el^2)$, where $l$ is the mean free path. In our
samples $B_{tr}\approx1$~T that roughly coincides with the field
at which the crossover from the one type of MR to the other takes
place. We conclude therefore that the strong MR observed at low
fields can be associated with the WL suppression in our samples
and that the MR observed at higher fields must be attributed
entirely to the e-e interaction effects \cite{Altshuler}.

  The longitudinal conductivity value is a sum of three components:
the classical Drude conductivity, the WL contribution and the e-e
interaction correction which is supposed to be independent of $B$
as long as $k_BT\tau/\hbar\ll1$. For the correct evaluation of the
interaction related correction at $B=0$~T, the knowledge of the
first two contributions to the conductivity is essential.
Unfortunately, in our case there is no direct means of knowing the
value of the Drude conductivity $\sigma_D$ because of a
considerable (up to $20\%$) variation of the zero field
conductivity with temperature. Nevertheless there exists an
empirical method that can be used for the evaluation of the all
three contributions to the conductivity at zero magnetic field.
This method has the advantage that one can forgo the usual
procedure of fitting the low field data with the theoretical
expressions for the WL magneto-resistance\cite{Hikami}, thus
eliminating a possible source of error at this stage.

  As a first step of this method the experimental longitudinal conductivity
is obtained by inverting the resistivity tensor using the data shown in
Fig.~\ref{Figure1}. The conductivity can be written as:

\begin{equation}
\label{Eq1}
\sigma_{xx}(T,B)=\frac{\sigma_D}{1+(\omega_c\tau)^2}+\Delta\sigma_{xx}^{WL}(T,B)+\Delta\sigma_{xx}^{ee}(T),
\end{equation}

  where $\omega_c$ is the cyclotron frequency, $\delta\sigma_{xx}^{WL}$ and
$\delta\sigma_{xx}^{ee}$ are the WL and e-e interaction
corrections respectively. The first term corresponds to the
classical $T$-independent Magneto-Conductivity (MC). The e-e
interaction correction, $B$-independent in the diffusive regime
\cite{Altshuler} is expected to become magnetic field dependent in
the opposite ballistic limit. The weak localization corrections
dominates at low fields but is suppressed at ($B>B_{tr}$).
Therefore, in the diffusive limit and for $B>>B_{tr}$ the shape of
the $\sigma$ vs $B$ dependence will be determined by the first
term in Eq.~\ref{Eq1} while the e-e interaction correction to the
conductivity should only result in a vertical shift of this
classical contribution. Indeed, experimentally we find that with
the WL completely suppressed at higher magnetic fields the MC
corresponding to different temperatures forms parallel vertically
shifted traces whose shape is given by the classical term in
Eq.~\ref{Eq1}. However, one can notice that at temperatures
$T>30$~K the shape of the curves begins to deviate slightly from
that of the low temperature traces. This change of shape may be
the consequence of the interaction correction becoming magnetic
field dependent at the crossover from the diffusive to the
ballistic regime.

  It is possible to determine the momentum relaxation time by fitting the
curves for $B>6$~T using the expression for the classical MC with
$\tau$ as a fitting parameter. This was done for all the
temperatures yielding the average value
$\tau=2.17\times10^{-14}$~s with a maximum deviation of $\approx
10\%$ . This value of $\tau$ corresponds to
$\sigma_D\approx6\times e^2/h$.\\ Next, in order to eliminate the
WL contribution at $B=0$ the term
$\frac{\sigma_D}{1+(\omega_c\tau)^2}+\Delta\sigma_{xx}^{ee}$ was
extrapolated for each of the curves down to $B=0$~T. Finally, to
obtain the value of the e-e interaction correction corresponding
to a given temperature, the Drude conductivity was subtracted from
the corresponding zero field conductivity value obtained at the
preceding step. Of course it is well to keep in mind that this
method is correct only as long as the e-e correction is
$B$-independent. That means that it is fully reliable only at low
temperatures where the MC traces are parallel. At $T>30$~K it will
give rise to an error increasing in proportion to the variation of
the shape of the curves. To remedy this we have also used an
alternative way to estimate the e-e interaction correction at high
$T$ which is as follows. The low temperature zero field WL
contribution determined at $T<30$~K with the method described
above was extrapolated to higher temperatures using the common
logarithmic law expression\cite{Lee}. The e-e correction was then
obtained by subtracting from the experimental zero field
conductivity the WL contribution obtained in this way together
with the Drude conductivity.

\begin{figure}[h]
\includegraphics[angle=-90,width=\columnwidth]{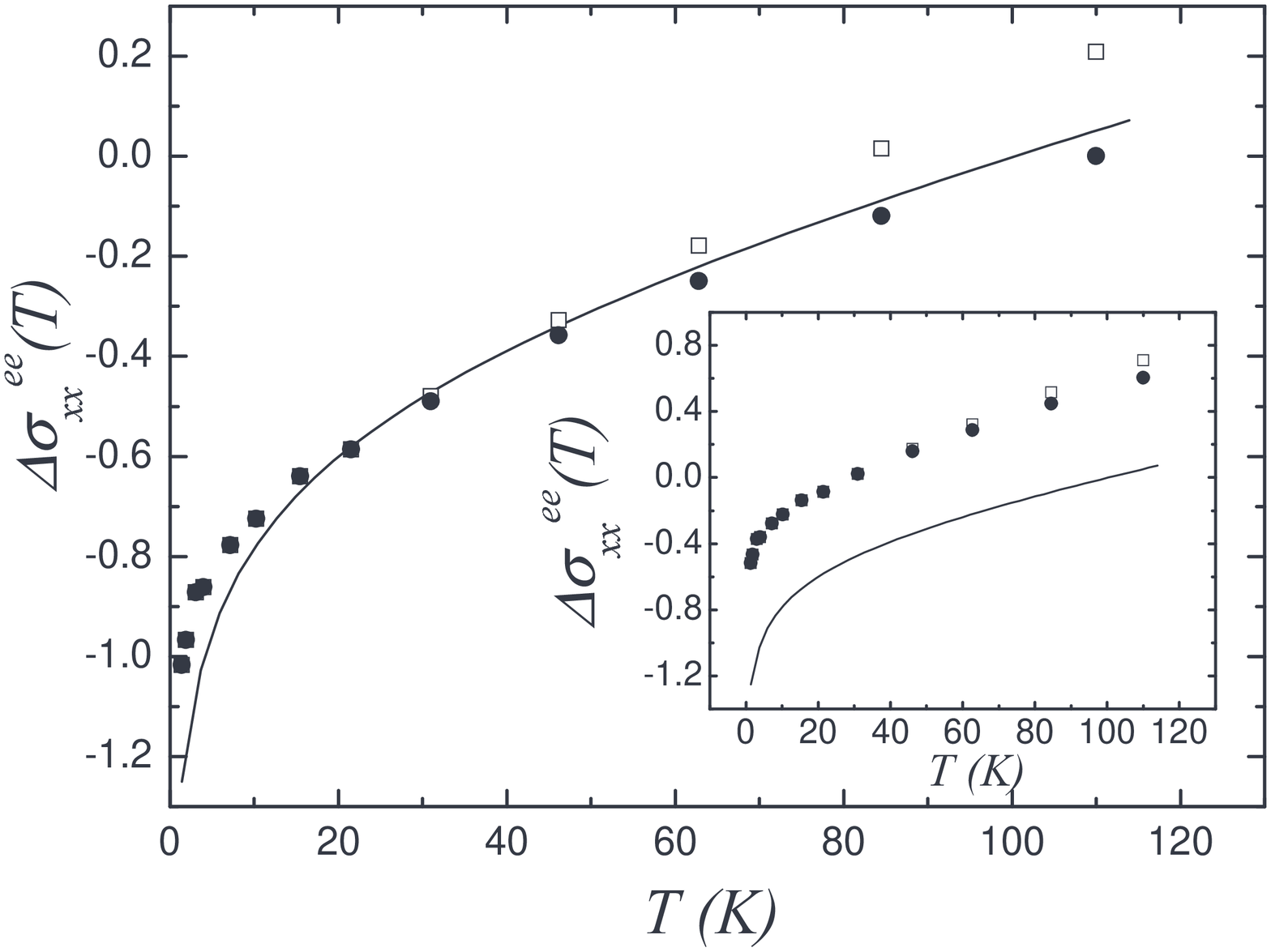}%
\caption{\label{Figure2} Experimental temperature dependence of
the e-e correction to conductivity obtained by the first method
(dots) and by the second method (opened squares), see the text.
They are compared to the model of Ref.~\onlinecite{Zala} (solid
line). The inset shows the correction obtained using the first
approximation of $\sigma_D$.}
\end{figure}

  The results obtained by these two different method are presented in the
insert to Fig.~\ref{Figure2}. We realize that none of these two
methods is fully accurate yet we suppose that the correct result
for the e-e interaction contribution must lie somewhere between
these two estimates.

  According to Ref.~\onlinecite{Zala} the e-e interaction correction to
conductivity is given by the following expressions:

\begin{equation}
\label{Eq2} \Delta\sigma_{xx}^{ee}=\delta\sigma_C+3\delta\sigma_T
\end{equation}
$$\delta\sigma_C=\frac{e^2}{\pi\hbar}\frac{k_{B}T\tau}{\hbar}\left[1-\frac{3}{8}
f(k_{B}T\tau/\hbar)\right]-
\frac{e^2}{2\pi^2\hbar}ln\left[\frac{\hbar}{k_{B}T\tau}\right];$$
is the charge channel correction and
$$\delta\sigma_T=\frac{F_{0}^{\sigma}}{\left[1+F_{0}^{\sigma}\right]}
\frac{e^2}{\pi\hbar}\frac{k_{B}T\tau}{\hbar}
\left[1-\frac{3}{8}t(k_{B}T\tau/\hbar;F_{0}^{\sigma})\right]$$
$$-\left[1-\frac{1}{F_{0}^{\sigma}}ln(1+F_{0}^{\sigma})\right]
\frac{e^2}{2\pi^2\hbar}ln\left[\frac{\hbar}{k_{B}T\tau}\right];$$

  is the correction in the triplet channel. The detailed expression of
$f(x)$ and $t(x)$ can be found in Ref.~\onlinecite{Zala}. It is
worth mentioning that for small $r_s$ the interaction constant
$F_0^\sigma$ is an analytical function of parameter $r_s$ (see
Ref.~\onlinecite{Zala}). In our calculations we used the value
$r_s=0.35$ corresponding to the electron density in our sample.

 Inset to Fig.~\ref{Figure2} shows the theoretical curve calculated for our
system parameters together with the experimental data points. As
can be seen there is a systematic shift of the experimental points
with respect to the theoretical curve. This shift can be
attributed to a not quite accurate evaluation of the Drude
conductivity. Indeed, a 10\% variation of $\sigma_D$ brings the
experimental data points closer to the corresponding theoretical
curve. A variation of this order of magnitude lies within the
experimental accuracy with which we determine $\sigma_D$ and only
weakly affect the shape of the theoretical curve.
Fig.~\ref{Figure2} shows the results obtained using
$\tau=2.33\times10^{-14}$~s ($\sigma_D=6.5\times e^2/h$). Thus a
reasonably good agreement for the entire temperature range which
also covers the intermediate regime is found.

  Note that contrary to the previous works
\cite{Coleridge,Shashkin,Proskuryakov,Kvon,Olshanetsky,Yasin} we
have used no fitting parameter. Moreover we find that using the
interaction constant $F_0^\sigma$ as a fitting parameter does not
result in a better agreement between theory and experiment.

\section{Hall effect}

  We now turn to the analysis of the Hall data presented in
Figure.~\ref{Figure1}b. According to Ref.~\onlinecite{Zala1} the
Hall resistivity may be written as:
\begin{equation}
\rho_{xy}=\rho_H^D+\delta\rho_{xy}^C+\delta\rho_{xy}^T \label{Eq3}
\end{equation}
where $\rho_H^D$ is the classical Hall resistivity and
$\delta\rho_{xy}^C,\delta\rho_{xy}^T$ are the corrections in the
charge and triplet channel. These corrections are given as follows:
\begin{equation}
\frac{\delta\rho_{xy}^C}{\rho_H^D}=\frac{2}{\pi}\frac{G_0}{\sigma_D}ln\left(1+\lambda\frac{\hbar}{k_BT\tau}\right)
\label{Eq4}
\end{equation}
$$\frac{\delta\rho_{xy}^T}{\rho_H^D}=\frac{6}{\pi}\frac{G_0}{\sigma_D}h(F_0^\sigma)ln\left(1+\lambda\frac{\hbar}{k_BT\tau}\right)$$

The detailed expression for $h(x)$ can be found in
Ref.~\onlinecite{Zala1}, $\lambda=\frac{11\pi}{192}$ and the value
of $\rho_H^D$ is obtained from the high temperature curves for
which
$\delta\rho_{xy}\rightarrow0$.\\

\begin{figure}[h]
\includegraphics[angle=-90,width=\columnwidth]{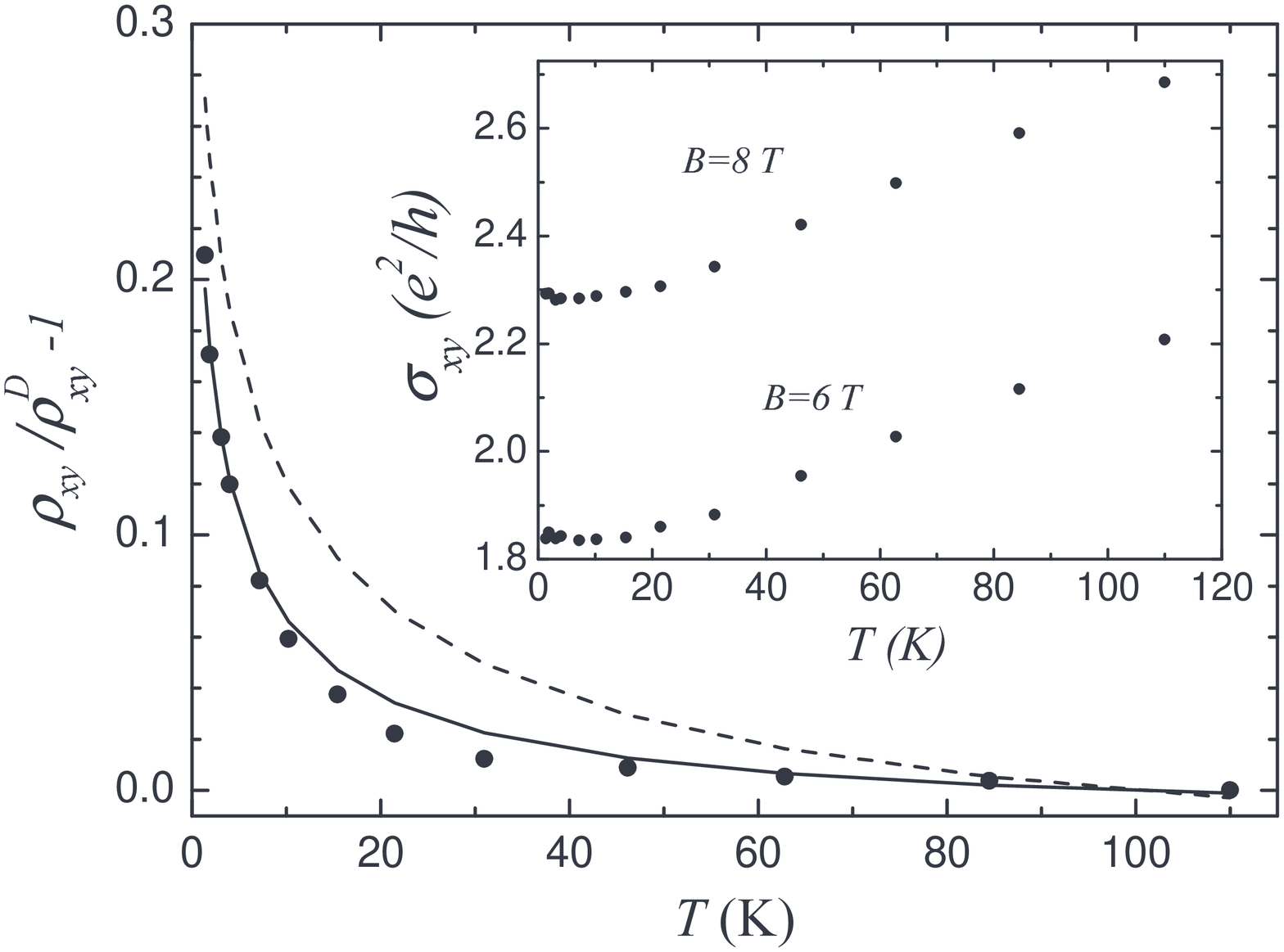}%
\caption{\label{Figure3} Temperature dependence of the Hall
coefficient (dots) compared to Eq.~\ref{Eq3} (dash line) and to
Eq.~\ref{Eq3} with $\lambda=\frac{3\pi}{192}$ (solid line). The
inset shows the transverse conductivity as function of $T$ for two
different values of magnetic fields.}
\end{figure}

  Therefore according to the theory of the e-e interaction \cite{Zala1} one should
observe a logarithmic temperature dependence of
$\rho_{xy}/\rho_{H}^{D}-1$ in the diffusive regime replaced by a
hyperbolic decrease $1/T$ at higher temperatures.
Figure~\ref{Figure3} shows how this prediction works in our case.

  A simple calculation (carried out without any attempt at fitting the
experiment) results in the dashed curve. In this calculation we
used the value
$F_0^\sigma\rightarrow\frac{1}{2}\frac{r_s}{r_s+\sqrt{2}}=-0.1$,
calculated with the expression recommended in the
theory\cite{Zala} for the weak interaction limit. On the whole
there is a qualitative agreement between theory and experiment
(black dots) but the quantitative agreement is lacking. Using
$F_0^\sigma$ as a fitting parameter does not improve the
agreement. Nevertheless we have found, that if the coefficient
$\lambda=\frac{11\pi}{192}$ is replaced by
$\lambda=\frac{3\pi}{192}$, then the theoretical curve (the solid
line) fits the experimental dependence quite well. This result
might be related to an anisotropy of electron scattering in the
sample which reduces the electron return probability and so
weakens the correction at low fields ($\omega_c\tau<<1$). The
reduction of the pre-factor $\lambda$ could just be the way in
which this anisotropy reveals itself since the correction is
proportional to $\lambda$ in the ballistic limit.

  Finally, in the inset to Figure.~\ref{Figure3} we show the experimental
data points for the transverse conductivity as a function of
temperature for two different values of magnetic field. In our
opinion these dependencies can serve as a good illustration for
the transition between the diffusive and ballistic interaction
regimes. Indeed, there should be no contribution of the
WL\cite{Altshuler} to the transverse conductivity tensor
component. Also, in the diffusive regime\cite{Altshuler}
$\Delta\sigma_{xy}^{ee}=0$. Thus, one would expect that
$\Delta\sigma_{xy}$ would be $T$-independent at low $T$-range
which is exactly what we see in the Inset. As for the intermediate
and ballistic regime, up to date there have been no predictions
concerning $\Delta\sigma_{xy}$. According to
Ref.~\onlinecite{Zala} the transition between the diffusive and
the ballistic regime occurs at $k_BT\tau/\hbar\approx0.1$
corresponding to $T\approx30$ K in our sample. One can see that at
about this temperature $\Delta\sigma_{xy}$ starts rapidly
increasing. One can conclude that the theoretical results for
$\Delta\sigma_{xy}$ in the diffusive limit are no longer valid in
the ballistic transport regime. To our knowledge this is the first
measurements of the $\sigma_{xy}$ temperature dependence in the
ballistic regime.

\section{Conclusion}

  In conclusion, we have observed the transition from the diffusive to the
ballistic regime in the weak interaction limit for the
longitudinal conductivity, the Hall coefficient, and the Hall
conductivity in a high density low mobility 2DEG. We find our
experimental results to be in a good qualitative agreement with
ZNA theory.
\begin{acknowledgments}
We are very grateful to A. Dmitriev, I. Gornyi, V. Tkachenko and
S. Studenikin for enlighting
discussions. This work was supported by PICS-RFBR (Grant No
1577.), RFBR (Grant No 02-02-16516), NATO, INTAS (Grant No
01-0014), programs "Physics and Technology of Nanostructures" of
the Russian ministry of Industry and Science and "Low dimensional
quantum structures" of RAS.
\end{acknowledgments}

\end{document}